\documentclass{aastex}

\usepackage{graphicx}
\usepackage{emulateapj5}
\usepackage{times}
\usepackage{psfig}



\newcommand{\lsim}{\mbox{\hspace{.2em}\raisebox{.5ex}{$<$}\hspace{-.8em}\raisebox{-.5ex}{$\sim$}\hspace{.2em}}}
\newcommand{\gsim}{\mbox{\hspace{.2em}\raisebox{.5ex}{$>$}\hspace{-.8em}\raisebox{-.5ex}{$\sim$}\hspace{.2em}}}

\def\chandra    {{\em Chandra}\/}

\def\xmm        {XMM-{\em Newton}\/}

\def\hst        {{\em HST}\/}

\def\ga         {{ESO~137-001}\/}

\begin{document}

\title{H$\alpha$ tail, intracluster HII regions and star-formation: ESO~137-001 in Abell 3627}

\author{
M.\ Sun,$^{\!}$
M.\ Donahue,$^{\!}$
G.\ M.\ Voit$^{\!}$
}

\affil{Department of Physics and Astronomy, MSU, East Lansing, MI 48824; sunm@pa.msu.edu}

\shorttitle{Star formation in the halo and the intracluster space}
\shortauthors{Sun et al.}

\begin{abstract}

We present the discovery of a 40 kpc H$\alpha$ tail and at least 29 emission-line
objects downstream of a star-forming galaxy \ga\ in the rich, nearby cluster A3627.
The galaxy is known to possess a dramatic 70 kpc X-ray tail.
The detected H$\alpha$ tail coincides positionally with the X-ray tail.
The H$\alpha$ emission in the galaxy is sharply truncated on the front
and the sides near the nucleus, indicating significant ram
pressure stripping. \ga\ is thus the first cluster late-type galaxy known
unambiguously with both an X-ray tail and an H$\alpha$ tail. The emission-line
objects are all
distributed downstream of the galaxy, with projected distance up to 39 kpc
from the galaxy. From the analysis on the H$\alpha_{\rm off}$ frame
and the estimate of the background emission-line objects, we conclude
that it is very likely all 29 emission-line objects are
HII regions in A3627. The high surface number density and luminosities
of these HII regions (up to 10$^{40}$ ergs s$^{-1}$) 
dwarf the previously known examples of isolated HII regions in clusters.
We suggest that star formation may proceed in the stripped ISM, in both
the galactic halo and intracluster space.
The total mass of formed stars in the stripped ISM of \ga\ may approach several
times 10$^{7}$ M$_{\odot}$. Therefore, stripping of the ISM
not only contributes to the ICM, but also adds to the intracluster
stellar light through subsequent star formation. 
The data also imply that \ga\ is in an active stage of transformation,
accompanied by the build-up of a central bulge and depletion of the ISM.

\end{abstract}


\section{Introduction}

The intracluster medium (ICM) has long been thought to play a vital role
in galaxy evolution in clusters (Gunn \& Gott 1972).
Ram pressure and turbulence/viscous
stripping by the ICM can efficiently remove cold galactic gas and may be
responsible for transforming disk galaxies with active star formation
into red S0 galaxies (e.g., Quilis et al. 2000). Stripping of the galactic
ISM in clusters has been extensively
examined in simulations (e.g., Schulz \& Struck 2001, SS01 hereafter; Roediger \&
Hensler 2005, RH05 hereafter), which show that stripping
has significant impacts on the properties of galaxies. In the process of
stripping, the galactic ISM is removed from the disk and the outer galactic
disk is truncated, while the inner disc is
compressed, accompanied by formation of numerous flocculent arms (SS01; RH05).
The galactic SFR is also modified. Compression of disk ISM may trigger prodigious
star formation (e.g., SS01; Bekki \& Couch 2003; RH05) in the first
$\sim 10^{8}$ yr of interaction, even though star formation activity
will eventually be suppressed as the galactic ISM is depleted. This initial
starburst may explain the blue ``Butcher-Oemler galaxies''
in $z\gsim$0.3 clusters (Butcher \& Oemler 1978).
The further evolution of the stripped ISM from the disk is not clear because
of uncertainties in the transport coefficients (conductivity and viscosity).
It has been
suggested that some stripped ISM may be able to cool to form stars because
the main ISM heating source, stellar UV radiation, is much weaker
in intracluster space (e.g., SS01). Therefore, ISM stripping in clusters is
important for both galaxy evolution and star formation in clusters.

Observational evidence of stripping in cluster late-type galaxies has only
begun to accumulate in recent years. Tails behind late-type galaxies have 
been observed in HI, H$\alpha$ and X-rays (e.g., Gavazzi et al. 2001;
Wang et al. 2004; Oosterloo \& van Gorkom 2005; Sun \& Vikhlinin 2005;
Yagi et al. 2007). Recently we found a long X-ray tail in a rich cluster
A3627, associated with a late-type galaxy \ga\ (Sun et al. 2006, S06 hereafter).
The narrowness and length of the
tail make it the most dramatic X-ray tail of a late-type galaxy to date.
It is also the only known X-ray tail from a
late-type galaxy in a rich cluster. There are only two other known
X-ray tails of late-type galaxies in clusters (C153 in A2125, Wang et al. 2004;
UGC~6697 in A1367, Sun \& Vikhlinin 2005). Both are in $\sim$ 3 keV clusters
that are more distant than A3627 (note $z=0.253$ for A2125).
We have also examined all \chandra\ and \xmm\ data for $T>$3 keV clusters
at z$<$0.1 (62 clusters in total; for the work on a subsample at z$<$0.05, see Sun et al. 2007).
No other significant stripped tails of late-type galaxies have been detected,
which implies that \ga's current stage is of short duration.

A3627 is a nearby rich cluster ($z$=0.01625, $\sigma_{\rm radial}$ = 925 km/s
and $kT$=6 keV) rivaling Coma and Perseus in mass and galaxy content
(Kraan-Korteweg et al. 1996; Woudt et al. 2007). \ga\ is a blue emission-line galaxy
(Woudt et al. 2004) that is only $\sim$ 200 kpc from the cluster's
X-ray peak on the plane of the sky. Its radial velocity (4630 km~s$^{-1}$)
is close to that of A3627 (4871$\pm$54 km~s$^{-1}$, Woudt et al. 2007).
S06 found a long narrow X-ray tail behind \ga, in both
\chandra\ and \xmm\ data (also shown in Fig. 1). The tail extends
to at least 70 kpc from
the galaxy with a length-to-width ratio of $\sim$ 10. The X-ray tail is
luminous ($\sim$ 10$^{41}$ ergs s$^{-1}$), with an X-ray gas mass
of $\sim 10^{9}$ M$_{\odot}$. S06 interpret the 
tail as the stripped ISM of \ga\ mixed with the hot ICM,
while this blue galaxy is being converted into a gas-poor galaxy.
The \chandra\ data also reveal three hard X-ray point sources
($L_{X} \sim 10^{40}$ ergs s$^{-1}$) along the tail (Fig. 1),
and the possibility of all of them being background AGN is $<$ 0.1\%.
Thus, we suggested that some of them may be ultra-luminous X-ray sources
(ULXs) born from active star formation in the tail.
Nevertheless, the optical properties of this interesting galaxy are poorly
known in the literature (e.g., Woudt et al. 2004; 2007).
In this paper, we present the results from our H$\alpha$ and optical broad-band
imaging observations of \ga\ with SOAR, and spectroscopic observations with
the CTIO 1.5m telescope.
We adopt a cluster redshift of 0.01625 for A3627 (Woudt et al. 2007),
which is a little larger than what we used in S06 (0.0157 from Kraan-Korteweg et al. 1996).
Assuming H$_{0}$ = 73 km s$^{-1}$ Mpc$^{-1}$, $\Omega$$_{\rm M}$=0.24,
and $\Omega_{\rm \Lambda}$=0.76, the luminosity distance is 67.6 Mpc,
and 1$''$=0.317 kpc. 

\section{Observations and Data Reduction}

\subsection{SOAR observations}

\ga\ was observed with the 4.1 m Southern Observatory for Astrophysical Research
(SOAR) telescope on Cerro Pachon on Aug. 28,
2006 (UT) and Mar. 15, 2007 (UT). Both nights were clear and photometric.
The observations were made with the SOAR 
Optical Imager (SOI), which covers a 5.26$'$ square field of view with two
CCDs (2.6$'$ width each, separated by an 8.1$''$ gap). In the 2006 run
($\sim 0.7''$ seeing, 1.23 - 1.31 airmass),
two 20-minute exposures were collected with the H$\alpha$ filter and five
20-second exposures were taken in the $I$ band. However, the telescope guider happened
to be south of the galaxy in two H$\alpha$ exposures, blocking any emission
26$''$ south of
the galactic nucleus. In the 2007 run ($\sim 0.9''$ seeing, 1.17 - 1.29
airmass), we took four 20-minute exposures in the H$\alpha$ band, two
10-minute exposures in another narrow band close to the H$\alpha$ band
(we call it the H$\alpha_{\rm off}$ band hereafter), seven 90-second exposures
in the $B$ band and four 30-second exposures in the $I$ band.
The H$\alpha$ filter used is the CTIO filter ID 6649-76 ($\lambda_{\rm cen}$
= 6650 \AA, FWHM=77 \AA) as the redshifted H$\alpha$ line of \ga\ is centered
at 6664 \AA. The H$\alpha_{\rm off}$ filter used is the CTIO filter ID
6520-76 ($\lambda_{\rm cen}$ = 6530 \AA, FWHM=71 \AA). The bandwidths
of the two narrow-band filters overlap little (at 6590 \AA, the transmission
of the 6649-76 filter is 4.8\%, while that of the 6520-76 filter is 2.6\%).
Both narrow-band filters are 2 inches $\times$ 2 inches, so the
non-vignetted field is only $\sim 3.6' \times 3.5'$. Dithers of
15$'' - 30''$ were made between exposures. During the second half
of 2006, the triplet lens of the focal reducer in SOI became partially
unbounded, which caused the photometry to be uncertain to $\sim$ 0.2 mag
from the field center to the corners.
The lens of SOI was repaired in the beginning of 2007 so the photometry
accuracy was recovered before our 2007 run.
In this work, we use only the 2007 data in the analysis.
Nevertheless, we examined the emission-line objects (see the definition and
discussion in $\S$3) in the 2006 data (which had better seeing). All 29 emission-line
objects selected from the 2007 data ($\S$3) are confirmed in the 2006 data,
although several of them are close to the guider in the 2006 data so their
fluxes are reduced.

Each image was reduced using the standard procedures with the IRAF
MSCRED package. The pixels were binned 2$\times2$, for a scale of
0.154$''$ per pixel. The fringing on the $I$ band images was subtracted
with the standard fringe frames. Dome flats were used. The US Naval
Observatory (USNO) A2.0 catalog was used for the WCS alignment, which
was done with WCSTools. Spectrophotometric standards are LTT~4364 and
EG~274 (Hamuy et al. 1994). Ten standard stars in two fields (PG~1047
and RU~149) on the $B, V, R$ and $I$ bands were also observed at airmass
of 1.1 - 1.8. \ga\ is near the Galactic plane ($l$=325.25 deg, $b$=-6.97 deg)
so the Galactic extinction is substantial. The Galactic extinction at the
$B$, $I$, H$\alpha$ and H$\alpha_{\rm off}$ bands is 0.89 mag, 0.40 mag,
0.554 mag and 0.568 mag respectively, derived from the Galactic
extinction law by Cardelli, Clayton \& Mathis (1989) for E(B-V) = 0.207 mag
(from NED). As \ga\ is not far from the Galactic center, the uncertainties
of Galactic extinction may not be small, but our conclusions of this paper
are hardly affected.

\subsection{CTIO spectroscopic observations}

Spectroscopic observations of \ga\ were conducted with the R-C
spectrograph on the 1.5m telescope in CTIO, operated by the Small and Moderate
Aperture Research Telescope System (SMARTS) Consortium. 
Two service observations were made on June 20 and 21, 2007, with the 47/Ib and
26/Ia grating setups respectively. The 26/Ia setup covers a wavelength range of
3660 - 5440 \AA\ with a dispersion of 1.5 \AA\ per pixel. The spectral resolution
is $\sim$ 4.3 \AA. The 47/Ib setup covers a wavelength range of 5652 - 6972 \AA\
with a dispersion of 1.1 \AA\ per pixel. The spectral resolution is $\sim$ 3.1
\AA. The slit is always aligned east-west with a length of $\sim 5'$ and a
width of 3$''$. The pixel scale on the CCD is 1.3$''$ per pixel and the best
instrumental spatial seeing for the R-C spectrograph is only $\sim 3.5''$.
For the 26/Ia observation (1.22-1.37 airmass), the slit was set across
\ga's center. Five exposures (20 minutes each) were made in the 26/Ia run
and we combined them.
The 47/Ib run on June 20, 2007 (1.60-1.91 airmass) was composed of three
exposures (25 minutes each). The slit was moved by several arcsec in the
north-south direction between exposures by mistake, although it was requested
in a fixed position the same as the one for the 26/Ia observations.
Thus, we analyzed the three exposures separately.
The pattern of the stellar spectra on the CCD (through the long
slit) allows us to recover the slit positions.
Spectroscopic standards were observed at each night, LTT~4364 for the 26/Ia run
and Feige~110 for the 47/Ib run. As these observations of the standards used
a 2$''$ slit, absolute flux calibration is impossible, but the analysis
of the emission line ratios should not be affected.

\section{Emission-line objects}
 
\subsection{Selection of emission-line objects}

With data in two narrow bands (H$\alpha$ and H$\alpha_{\rm off}$) and two broad
bands ($B$ and $I$), we derived color-color and color-magnitude relations
for sources in the field covered by all data (3.42$'\times3.88'$ around
\ga\ shown as the dotted box in Fig. 1b). SExtractor 2.5.0 was used to measure photometry
and color. The H$\alpha$ and H$\alpha_{\rm off}$ magnitudes are AB
magnitudes defined in Hamuy et al. (1994) and are calibrated with the data
of EG 274 and LTT 4364. Galactic extinction has been corrected in all bands.
In the H$\alpha$ - H$\alpha_{\rm off}$ vs. H$\alpha - I$ map (Fig. 2),
over 97\% of 1708 sources in the field are Galactic stars
with little excess or decrement emission in the H$\alpha$ filter.
The H$\alpha$ - H$\alpha_{\rm off}$ histogram peaks at $\sim$ -0.02 mag and
the typical 1 $\sigma$ dispersion is $\sim$ 0.15 mag.
The H$\alpha - I$ histogram peaks at $\sim$ 0.46 mag and is skewed to
higher color values. The 1 $\sigma$ dispersion on the lower color side is $\sim$ 0.22 mag.
Besides these stars, there are about 30 sources that are especially bright
in the H$\alpha$ band and occupy a different region in the color - color map.
In this work, we selected 29 emission-line objects with the following conservative
criteria: H$\alpha$ - H$\alpha_{\rm off} < -0.7$ mag and H$\alpha - I < -0.2$ mag.
These emission-line objects have equivalent widths (EW) of 84 - 758 \AA\
(we adopted the definition of EW in Gavazzi et al. 2006).
Their positions, as shown in Fig. 1b, are all downstream of the galaxy and in
or around the X-ray tail. If we relax the criteria to
H$\alpha$ - H$\alpha_{\rm off} < -0.5$ mag and H$\alpha - I < 0.25$ mag,
six additional sources are selected and they are also in or around the X-ray
tail (Fig. 1b). 

The H$\alpha$ mag vs. H$\alpha - I$ relation is also shown in Fig. 2. After
correcting the atmospheric and Galactic extinction, the fluxes of these
emission-line objects range from 1.5$\times10^{-16} - 4.0\times10^{-14}$
ergs s$^{-1}$ cm$^{-2}$ (no correction for intrinsic extinction). We also
derived the $B - I$ color for sources in the field. As some emission-line
objects are not detected in the $B$ and $I$ bands, $B - I$ color can be
constrained only for 17 emission-line objects (Fig. 2). The emission-line objects
are generally bluer than the galaxy (without correction for intrinsic
extinction) and Galactic stars. The $B - I$ colors of Galactic stars
span from M type (3.6 - 5.1 mag) to A type (0 - 0.4 mag). As the Galactic reddening
(0.49 mag) was applied on all sources, the actual separation between $B - I$
colors of emission-line objects and Galactic stars will be wider if zero or smaller
Galactic extinction applies on stars and some amount of intrinsic extinction
is applied for emission-line objects. Therefore, these emission-line objects
are blue and a small amount of intrinsic extinction can drive their colors
to those of pure O/B star clusters.

\subsection{What are they?}

What is the nature of these emission-line objects? 
They cannot be Galactic H$\alpha$ emitters as the required velocity
(2600 km/s - 5300 km/s) is too high. The other emission line close to the
H$\alpha$ filter band is [SII] 6716\AA\ (but generally with small EW).
However, a velocity of $<$ -1600 km/s is required to put the [SII] line in the
H$\alpha$ filter band, which is also too large for Galactic objects. Thus,
these emission-line objects are extragalactic.
Besides being H$\alpha$ emitters in A3627, these objects could also be
[OIII] $\lambda$5007 emitters at $z \sim 0.33$, H$\beta$ emitters at $z \sim 0.37$,
[OII] $\lambda$3727 emitters at $z \sim 0.78$
and Ly$\alpha$ emitters at $z \sim 4.47$.
We have examined the expected number density of these background objects
in the following two ways.

First, we can examine the H$\alpha_{\rm off}$ frame in the same way as we did
for the H$\alpha$ frame to select emission-line objects in the H$\alpha_{\rm off}$
band (or sources with large H$\alpha$ - H$\alpha_{\rm off}$ color).
Both narrow-band filters have very similar transmission and band width. The
difference in their central wavelength is small. Thus, we expect that the number
of background emission-line objects detected in the two narrow bands should be similar.
H$\alpha_{\rm off}$ - H$\alpha$ vs. H$\alpha_{\rm off} - I$ and H$\alpha_{\rm off}$
vs. H$\alpha_{\rm off} - I$ maps (shown in Fig. 2) are derived for sources
detected in the H$\alpha_{\rm off}$ band. The same criteria used to select
emission-line objects in the H$\alpha$ band were applied (with just
the change from H$\alpha$ magnitude to H$\alpha_{\rm off}$ magnitude).
No emission-line objects have been found in the H$\alpha_{\rm off}$ band,
even though the loose criteria are applied (H$\alpha_{\rm off}$ - H$\alpha <$
-0.5 mag and H$\alpha_{\rm off} - I <$ 0.25 mag).  In fact, the H$\alpha_{\rm off}$
exposure is deep enough to detect the faintest emission-line object in the
H$\alpha$ band, if a similar object is present in the H$\alpha_{\rm off}$ frame.
This single self-test already demonstrates that the expected
number of background emission-line objects is very small.

Second, we examined the number density of emission-line objects from the
Large Area Ly$\alpha$ survey at $z \sim 4.5$ (Malhotra \& Rhoads 2002) as
they used the narrow-band filters with similar central wavelength and
bandwidth ($\sim$ 80\AA) to ours.
Their covered field is 0.72 deg$^{2}$ and the achieved sensitivity is
about 5 times deeper than ours. In three non-overlapping H$\alpha$ filters
(the central one has the same central wavelength as our H$\alpha$ filter),
157 Ly$\alpha$ candidates (with EW $> 80$\AA\ and without $B$ band detection)
were detected. The size of our field is 3.42$'\times3.88'$.
Thus, the expected number of emission-line objects selected in Malhotra \& Rhoads
(2002) is $\sim$ 0.02 in our field (adjusted to our flux limit by
assuming a power index of -1.6 for the luminosity function, Malhotra \& Rhoads
2002). Cortese et al. (2004) examined $\sim$ 2.5 deg$^{2}$ H$\alpha$ frames of
the Virgo cluster ($\lambda_{\rm cen} = 6574$ \AA, $\Delta \lambda$  = 95 \AA)
obtained at the Isaac Newton Telescope.
Six emission-line objects were detected (for a H$\alpha$ flux limit of about
6 times shallower than the faintest emission-line object in our sample) and their
follow-up spectroscopic observations showed that five of them were [OIII] $\lambda$5007
emitters at $z \sim 0.31$. If we take their number as the number density of
$z \sim 0.33$ [OIII] $\lambda$5007 emitters in the high luminosity and apply a
luminosity function with a power index of -1.4 (e.g., Pascual et al. 2001), the
expected number of [OIII] emitters in our whole field is $\sim$ 0.09.
For our line flux limit ($\gsim 10^{16}$ ergs s$^{-1}$ cm$^{-2}$), the combined
number density of the H$\beta$ and the [OII] $\lambda$3727 emitters is at most
comparable to that of [OIII] $\lambda$5007 emitters (e.g., Pascual et al. 2001),
and these two lines generally don't have high EW.

Therefore, the expected number of background emission-line sources is
much smaller than one in our 3.42$'\times3.88'$ field and it is very
likely that {\it all 29 emission-line objects are H$\alpha$ emitters associated
with A3627}. In fact, this conclusion is even much stronger when we
consider the spatial concentration of emission-line objects immediately
downstream of \ga\ and around the X-ray tail. {\it These 29 + 6 sources are
not randomly distributed in the field and relative to the galaxy.}
Twenty-five emission-line objects cluster in a 40$''\times60''$ box
region immediately downstream of \ga, while the area of our total field
is 20 times larger. 
In $\S$4.1, we further show that seven emission-line objects are caught
in one or more CTIO 1.5m slit spectra. The spectra strongly support the
argument of HII regions ($\S$4.1), at least for these seven emission-line
objects. As H$\alpha$ emitters in A3627, they are too luminous
to be intracluster planetary nebulae. The H$\alpha$ luminosity of the
faintest emission-line object in our sample is $>$ 7.2 times larger
(without intrinsic extinction) than the [OIII] $\lambda$5007 luminosity
of the most luminous intracluster planetary nebulae in the Virgo cluster
(Ciardullo et al. 1998) and the H$\alpha$ line of a planetary nebula is
generally several times fainter than the [OIII] line. Thus, we conclude
that they are HII regions most likely associated with \ga\ and its tail
in A3627. Hereafter, we simply refer these emission-line objects as HII regions.

\subsection{Properties of the HII regions}

The properties of these HII regions and the embedded star clusters can be
estimated. The intrinsic extinction is unknown. Gerhard
et al. (2002) and Cortese et al. (2004) measured $\sim$ 1 mag intrinsic
extinction for two isolated HII regions in the Virgo cluster. We simply
adopt this value for all HII regions. The [NII] $\lambda$6548 and
$\lambda$6584 lines are in the H$\alpha$ filter band. The [NII]
$\lambda$6584 line is close to the wing (transmission there is 65\% of
the value in the center). Gerhard et al. (2002) and Cortese et al. (2004)
measured H$\alpha$ / (H$\alpha$ + [NII] $\lambda$6548 + [NII]
$\lambda$6584) $\sim$ 0.81 for two isolated HII regions in the Virgo
cluster. This fraction is assumed in our analysis. The derived H$\alpha$
luminosities of these HII regions are plotted in Fig. 3 versus their
distance to \ga's nucleus. The cumulative luminosity function of these
HII regions, N ($>$ log $L) \propto L^{-\alpha}$, has a slope of
0.6$\pm$0.1 at $L_{H\alpha} > 10^{38.3}$ ergs s$^{-1}$. With the scaling
relation derived by Kennicutt (1998), SFR (M$_{\odot}$/yr) =
$L_{H\alpha} / (1.26\times10^{41}$ ergs s$^{-1}$), the SFR in these HII
regions ranges from 0.0008 to 0.17 M$_{\odot}$/yr, with a total SFR of
0.59 M$_{\odot}$/yr for 29 HII regions selected. Assuming an electron
temperature of 10$^{4}$ K and case B of nebular theory,
the number of ionizing photons $Q$(H) ranges from
9.5$\times10^{49}$ to 2.1$\times10^{52}$ s$^{-1}$.

We can apply the Starburst99 model (Leitherer et al. 1999) to estimate the
age and the total mass of the starbursts in these HII regions.
The age of the starburst can be estimated from the ratio $Q$(H) / $L_{B}$, or
the ratio $Q$(H) / $L_{I}$ (e.g., Gerhard et al. 2002), or the H$\alpha$
equivalent width --- EW (H$\alpha$). The first two estimates may only provide
lower limits on the age as the intrinsic extinction of the HII regions in the
$B$ and $I$ bands is unknown. EW (H$\alpha$) is not affected by intrinsic
extinction, although the uncertainties are generally large and there are
only lower limits in many cases. We selected eight representative HII regions
(marked ELO1 to 8 in Fig. 1c and 1d) to demonstrate the range of the properties
of these 29 objects. ELO1 is the brightest and the only resolved one (with a
radius of $\sim$ 0.6 kpc). ELO2 and 3 are the next two most luminous ones.
ELO4 is the one with the lowest EW (H$\alpha$). ELO5 is only $\sim 0.4''$
offset from the brightest \chandra\ hard X-ray point source in the tail (P1 in
S06). ELO6 is the faintest emission-line object, while ELO7 and 8 are most
distant from \ga's nucleus in projection (30 - 39 kpc). 
A metallicity of 0.4 solar is assumed, which is similar to those of two HII regions
studied by Gerhard et al. (2002) and Cortese et al. (2004). We assume instantaneous
star formation with a Salpeter IMF, $M_{\rm up} = 100 M_{\odot}$ and
$M_{\rm low} = 1 M_{\odot}$. The results are listed in Table 1.
The age estimated from $Q$(H) / $L_{B}$ is always the smallest, which implies
intrinsic extinction in HII regions. The age is always less than $\sim$ 7 Myr,
which is typical for bright HII regions (e.g., Gerhard et al. 2002;
Mendes de Oliveira et al. 2004). The estimated age from EW (H$\alpha$) is not
sensitive to the assumed metallicity for EW (H$\alpha$) $>$ 400 \AA\ HII regions.
Even for low EW (H$\alpha$) regions, the uncertainty of age is not very big.
For example, the predicted age of ELO4 from EW (H$\alpha$) is 6.7, 6.3, 6.5, 8.3,
and 11.5 Myr for assumed metallicity of 2, 1, 0.4, 0.2, 0.05 solar respectively.
It is clear that these HII regions are all young ($\lsim$ 10 Myr).
The derived total mass ($\sim 10^{3}$ M$_{\odot}$ --- $\sim 4\times10^{6}$ M$_{\odot}$)
is sensitive to the age of the starburst, as shown for two HII regions with similar
H$\alpha$ luminosities (ELO6 and 8, Table 1). The older the HII region is, the
more massive it is. Nevertheless, the brightest
5 - 10 HII regions host star clusters with total mass of $> 10^{5}$ M$_{\odot}$,
or super star clusters (10$^{5}$ - 10$^{8}$ M$_{\odot}$, O'Connell 2004).
The total mass in the 29 putative HII regions is about 10$^{7}$ M$_{\odot}$, with
at least a factor of four uncertainty because of the unknown extinction and
uncertain age.

The H$\alpha$ emission of HII regions fades rapidly for a single starburst and
will be much fainter after the initial 10 Myr. The radial velocity difference
between \ga\ and A3627 is only 214 km/s, which implies that \ga's infalling is
almost in the plane of the sky. If we assume a velocity of \ga\ equal to the velocity
dispersion of the cluster (1600 km/s), \ga\ travels only $\sim$ 16 kpc in 10 Myr.
It is clear that there are no very luminous HII regions beyond 15 kpc from the
nucleus (Fig. 3). We also examined the H$\alpha$ EW of the HII regions with
offset from the galaxy. There is no clear relation found as 13 sources have
only lower limits. Within 8 kpc from the nucleus, the H$\alpha$ EW span a big
range from 77 \AA\ to 614\AA. In any case, the three HII regions at 29 - 39 kpc
projected from the nucleus must have had star formation happening in the last
$<$ 10 Myr, at least 10 - 15 Myr after the gas was removed from \ga. Thus, the
star formation in these HII regions does not always start immediately after
they are displaced from the galactic disk, if star formation in these HII
regions is indeed a single burst. It is likely that more HII regions downstream
of \ga\ are yet to form and many HII regions may have already faded in or around
the tail. We discuss the population of the HII regions more in $\S$6.

\section{The optical properties of \ga}

\subsection{The spectroscopic properties}

The spectroscopic properties of the central emission nebula of \ga\ are
examined with the CTIO 1.5m spectra. The slit positions are recovered from the
stellar spectra on slits, and they are shown in Fig. 4.
The \#1 slit of the 47/Ib observations misses the central emission nebula of
the galaxy. The stellar spectrum at this offset position is too faint to
be studied with this single exposure, while a nearby star (close to
source ``a'' in Fig. 4, see also Fig. 1) is several times brighter.
The \#3 slit of the 47/Ib observations covers the H$\alpha$ peak of \ga\
and the resulting spectrum is the brightest.
The spectra of \ga's central part are shown in Fig. 5,
from the combined 26/Ia exposures and the \#3 47/Ib exposure. 
The measured velocity with RVSAO is 4667$\pm$135 km/s from the 26/Ia spectrum and
4640$\pm$20 km/s from the \#3 47/Ib spectrum, which is consistent with
4630$\pm$58 km/s measured by Woudt, Kraan-Korteweg \& Fairall (1999). 
As shown in Fig. 5, many line ratios
can be determined. However, the 26/Ia and 47/Ib spectra
were taken at different nights and at different positions. It is also impossible
to do absolute flux calibration for these spectra. Therefore,
we restrict our line ratio analysis only
to lines in the same wavelength ranges (26/Ia or 47/Ib), which makes
the usual method to constrain the intrinsic reddening with the
H$\alpha$ / H$\beta$ ratio impossible to be applied.
Instead, we use the H$\gamma$ / H$\beta$ ratio to constrain the intrinsic reddening.
The theoretical value for pure recombination is taken from Osterbrock (1989),
0.468 for 10$^{4}$ K
gas (case B) at $n_{e} = 10^{2}$ cm$^{-3}$. The line ratio measured is 0.35,
after the correction for the Galactic reddening. Using the Galactic extinction
curve from Cardelli et al. (1989) and assuming $R_{V} = A_{V} / E(B-V)$ = 3.1,
we derive $A_{V}$ = 1.7 mag for the intrinsic extinction along \ga's core.
The corresponding intrinsic extinction for the
H$\alpha$ line is 1.4 mag, which is consistent with the typical extinction
of the H$\alpha$ line for nearby spirals (0.5 - 1.8 mag, e.g., Kennicutt 1983)
and the likely near edge-on orientation of the galaxy (see $\S$5).
Therefore, this amount of intrinsic extinction has been applied to
the line ratio analysis in this paper.

The gas properties can be determined with the emission line ratios.
We derived various line ratios from the combined 26/Ia spectrum and the \#3
47/Ib spectrum as their emission comes from similar regions of the galaxy (Fig. 4).
We caution that the flux calibration at the blue end of the spectrum is more
vulnerable to the flat field correction so the error of the [OII] flux is bigger
than those of other strong emission lines.
The [SII] $\lambda$6716 / [SII] $\lambda$6731 ratio is $\sim$ 1.38, which is
typical for HII regions and is comparable to the low-density limit of 1.35
(van Zee et al. 1998). The strong [OII] line compared to the [OIII] lines
suggests a low ionization parameter of the gas. We derived:
log([OI] $\lambda$6300 / H$\alpha) \lsim$ -1.65, log([OIII] $\lambda$5007 / [OII])
= -1.39, log([NII] $\lambda$6584 / H$\alpha$) = -0.42, log([SII] $\lambda\lambda$6716,
6731 / H$\alpha$) = 0.39 and log([OIII] $\lambda$5007 / H$\beta$) = -0.51.
All these ratios indicate that the central emission nebula of \ga\ resembles a
typical giant HII region (Fig. 4 and 5 of Kewley et al. 2006).
The low [OI] $\lambda$6300 / H$\alpha$ ratio implies that any central
AGN, if existed, is very weak, which is consistent with the X-ray non-detection
of the nucleus (a 3$\sigma$ limit of 5$\times10^{39}$ ergs s$^{-1}$ in the
0.5 - 10 keV band).

The gas metallicity can be estimated from several emission line ratios 
(e.g., Kewley \& Dopita 2002), but these line ratios also depend on
the ionization parameter. The ionization parameter can be estimated from
the [OIII] $\lambda$5007 / [OII] ratio, 0.041, which implies an
ionization parameter ($q$) of $\lsim 10^{7}$ cm/s for gas metallicity of $<$
2.0 solar (Kewley \& Dopita 2002). We then estimate the gas metallicity
from the following line ratios for $q \lsim 10^{7}$ cm/s (Kewley \& Dopita 2002):

\begin{enumerate}
\item log([NII] $\lambda$6584 / [SII] $\lambda\lambda$6716, 6731) = 0.015,
which implies log(O/H)+12 = 8.75 - 8.95.
\item log([OII] $\lambda$3727 + [OIII] $\lambda\lambda$4959, 5007 / H$\beta$)
(or R$_{23}$) = 0.90, which implies log(O/H)+12 = 8.05 - 8.20 or $\sim$ 8.65.
\item log([NII] $\lambda$6584 / H$\alpha$) = -0.44, which implies
log(O/H)+12 $\sim$ 8.73 or 9.35.
\item log([NII] $\lambda$6584 / [OII]) = -0.84, which implies log(O/H)+12 =
8.62-8.75.
\end{enumerate}

As the line fluxes of [NII] and [OII] come from different observations, their
ratio is estimated by multiplying these three line ratios,
[NII] $\lambda$6584 / H$\alpha$,
H$\alpha$ / H$\beta$ and H$\beta$ / [OII]. For the H$\alpha$ / H$\beta$ ratio,
the theoretical value of 2.86 is used for case B recombination at 10$^{4}$ K
and $n_{e} \sim 100$ cm$^{-3}$ (Osterbrock 1989).
In spite of uncertainties, we find that log(O/H)+12 $\approx$ 8.7 (0.6 solar)
is consistent with all estimates.

As shown in Fig. 4, seven emission-line objects are also in one or more slit
spectra. A large part of ELO1's emission is in the 26/Ia spectra. However,
the best instrumental spatial seeing for the spectrograph is $\sim 3.5''$ (or 2.7
pixels). Since both ELO1 and the
H$\alpha$ core of \ga\ are extended (Fig. 1) and their peaks are only 4.6$''$
away, their lines will be blended in the spectra.
Nevertheless, we indeed observe extension of the H$\beta$, [OII] $\lambda$3727
and [OIII] $\lambda$5007 emission towards the direction of ELO1 (or upward on
the CCD plane, Fig. 4). The scale of the
extension is consistent with the position of ELO1.

The other six emission-line objects only appear as a single line in each
spectrum, because their continua are very weak. Although a single line 
detection does not determine their redshifts, the following arguments 
support that at least the bright ones among them are HII regions.
First, if we stack all three 2D spectra shown in Fig. 4, we detect significant [NII]
$\lambda\lambda$6548, 6584 emission at $\sim 5 \sigma$.
Second, the centroids of these lines are only $<$ 3\AA\ (or 137 km/s) higher than the
H$\alpha$ line of the galaxy.
Third, if the detected line is [OIII] $\lambda$5007 or H$\beta$ at higher
redshifts, we expect significant
H$\beta$ or [OIII] 5007 line also in the 47/Ib spectra from the usual
[OIII] 5007 / H$\beta$ ratio (Kewley et al. 2006). However, they are
not detected in either a single exposure or the stacked exposure.
Good quality spectra of these emission-line objects would require a big
telescope.

\subsection{The H$\alpha$ and broad-band images}

The photometric properties of \ga\ are also studied.
As shown in Fig. 1, the galaxy may have a distorted inner disk or bar within 1 kpc
radius. A dust lane is also significantly detected in all four bands at 0.5 kpc
to the south of the nucleus. However, detail around the nucleus can only be
obtained with \hst\ imaging. The morphological type of \ga\ (e.g., from Sb to Sd)
is also unclear.
Downstream in the $B$ and $I$ band images, some substructures are detected
to $\sim$ 6 kpc from the nucleus, including two blue streams (Fig. 1g).
Interestingly, these structures all have nearby emission-line objects.
Arm-like features are also detected at the north ($\sim 35''$ from the nucleus)
and the south ($\sim 26''$ from the nucleus).
To quantitatively examine the optical light distribution of \ga, we measured
surface brightness profiles along the minor axis and the major axis (Fig. 6).
Although we estimated the intrinsic reddening around the galactic center in
$\S$4.1, the intrinsic reddening outside the H$\alpha$ emission nebula is
unknown and may be smaller. Thus, we elect not to correct for the intrinsic
extinction for these profiles.
The H$\alpha$ emission is very asymmetric along the minor axis because of the
HII regions and the H$\alpha$ tail downstream of the galaxy.
The net H$\alpha$ emission is truncated sharply upstream at $\sim 0.9$ kpc
from the nucleus. The H$\alpha$ emission is much more symmetric along the
major axis, and the net H$\alpha$ emission is truncated sharply at $\sim 1.5$ kpc
from the nucleus. In the optical, the galaxy is composed of at least two components
(Fig. 6). The bright inner component extends to $\sim$ 1.9 kpc in radius along
the major axis and to $\sim$ 1.2 kpc in radius along the minor axis. Almost
all H$\alpha$ emission is within the inner component. 
Within the central 1.5 kpc radius, the $B$ and $I$ band light distributions
have multiple peaks.
The outer component can be fitted with an exponential profile and the derived
scale height is $\sim 6''$ along the minor axis and $\sim 12''$ along the major axis
(see the caption of Fig. 6). The light distribution is a little more extended
in the north, compared to that in the south (Fig. 6). 

We also measured the
light profiles in elliptical annuli centered on the H$\alpha$ peak (see Fig. 6).
Again, the light profiles at $a = 7'' - 32''$ ($a$: the semi-major
axis) can be fitted with an exponential profile with a scale height of $\sim 9.5''$,
while fits with the de Vaucouleurs 1/4 law overestimate the emission at $a > 25''$.
Although the measured light profiles (without correction on the intrinsic
reddening) imply little change of the $B - I$ color with $a$, the known
intrinsic reddening in the central H$\alpha$ nebula implies that the
$B - I$ color is 1.2 mag bluer there. This blue and bright core may be
a central bulge in formation. 
The orientation of \ga's disk can be estimated from the classical Hubble
formula\footnote{http://leda.univ-lyon1.fr/leda/param/incl.html}.
Assuming an axis ratio of $\sim$ 2 and a morphological type from
Sb to Sd, the estimated angle between the line of sight and the disk plane
is 26 - 29 deg. Thus, the putative disk is viewed close to edge-on.

We also derived the total $B$ and $I$ band magnitudes of \ga\ with the light profiles
derived in elliptical annuli (Fig. 6). The total $B$ band magnitude measured
within a 20$''\times40''$ ellipse is 14.31 $\pm$ 0.08 mag (or $\sim$ 2/3 $L*$
in the $B$ band), after correcting
the atmospheric and Galactic extinction. This value can be compared with
the $B$ magnitude listed on HyperLeda, 14.05 $\pm$ 0.23 mag from the ESO survey.
Our images are deeper and allow much better masking of Galactic stars as
some of them are very close to the nucleus (Fig. 1).
The total $I$ band magnitude measured in the same aperture is 13.20 $\pm$ 0.07 mag,
after correcting the atmospheric and Galactic extinction.
The half-light size of the galaxy is also estimated from the light profiles
measured in elliptical annuli: 14.5$''$ (or 4.45 kpc) semi-major axis in the $B$ band
and 14.0$''$ (or 4.30 kpc) semi-major axis in the $I$ band.
The 2MASS $K_{s}$ total magnitude is 12.163 mag, after correction for the
Galactic extinction (0.076 mag).
The intrinsic reddening measured for the H$\alpha$ emission nebula would imply
a corrected $B-K_{s}$ color of $\sim$ 0, although the real averaged value
should be higher as the regions outside of the H$\alpha$ emission nebula may
have smaller intrinsic extinction. In any case, the $B-K_{s}$ color of \ga\
is much bluer than that of a typical late-type galaxy ($\sim$ 2 - 3.5, Jarrett 2000).
The galaxy is way off the red-sequence of galaxies in clusters and is in the
so-called ``blue cloud'' in the cluster color-magnitude relation. 
The $K_{s}$ band absolute magnitude is -21.97 mag.
Bell et al. (2003) determined the local $K_{s}$ band luminosity function
and found an absolute magnitude of -23.97 mag for an $L*$ galaxy (with early-type
and late-type galaxies mixed, adjusted to the cosmology we used).
Therefore, \ga\ is a 0.16 $L*$ galaxy in the $K_{s}$ band.
The total stellar mass of \ga\ can also be estimated.
The mass-to-light ratios in the $I$ and $K_{s}$ bands can be estimated from
the relations derived in Bell \& De Jong (2001): 
$M / L_{I}$ = 0.59 $(M / L_{I})_{\odot}$, $M / L_{Ks}$ = 0.35 $(M / L_{Ks})_{\odot}$
(using the function for the $K$ band) for $B-I$ = 1.1 mag.
The resulting stellar mass of \ga\ is 4.9$\times10^{9}$ M$_{\odot}$ from $L_{I}$
and 4.6$\times10^{9}$ M$_{\odot}$ from $L_{Ks}$.
The total stellar mass will be smaller if a correction for the intrinsic reddening
is made. For example, for an intrinsic reddening of 0.5 mag for the $B-I$ color,
the total stellar mass of \ga\ will be one third smaller.
Bell et al. (2003) also determined the $M_{*}$ of the local stellar mass function,
$M_{*} = 8.0\times10^{10}$ M$_{\odot}$.
Therefore, though \ga\ is not a small galaxy, it is blue and has a low stellar
mass ($\sim 0.05 M*$).

The net H$\alpha$ emission of \ga\ was also studied. The continuum emission
in the H$\alpha$ image is determined from the H$\alpha_{\rm off}$ frame
and the scaling factor was determined to account for all emission of
stars in the H$\alpha$ image (or H$\alpha$ - H$\alpha_{\rm off}$ = 0,
Fig. 2). The net H$\alpha$ image is shown in Fig. 1c and 1d.
The net H$\alpha$ flux is also measured. The calibration sources
LTT~4364 and EG~274 have nearly flat spectra within the H$\alpha$ filter,
while the net narrow-band image of \ga\ is composed of three narrow lines.
We can correct for this spectral difference and subtract the [NII] flux
with the CTIO spectrum ($\S$4.1) and the transmission curve of the
H$\alpha$ filter. The H$\alpha$ and the [NII] $\lambda\lambda$6548,
6584 lines can be approximated by three Gaussians with $\sigma$ of $\sim$ 3\AA.
Their flux ratios are known from the CTIO data. 
Thus, the conversion factor from the count rate to flux can be derived.
The flux is first measured in an elliptical aperture with 3$''$ (semi-minor axis)
$\times 5''$ (semi-major axis) centered on the H$\alpha$ peak (Fig. 6), which
encloses almost all emission of the central nebula.
The H$\alpha$ EW is 50 \AA, which can be compared with the H$\alpha$ EW derived
from the CTIO spectra (26 \AA\ from the slit position \#2, 63 \AA\ from the slit
position \#3).
With an intrinsic extinction of 1.4 mag ($\S$4.1), the flux is 2.7$\times10^{-13}$
ergs s$^{-1}$ cm$^{-2}$ and the luminosity is 1.5$\times10^{41}$ ergs s$^{-1}$.
The total H$\alpha$ luminosity of \ga, measured in a boxy aperture (20$''\times26''$,
shown in Fig. 1d, including HII regions ELO1 and several others), is
2.5$\times10^{41}$ ergs s$^{-1}$ (still assuming
1.4 mag intrinsic extinction). With the scaling relation by Kennicutt (1998),
the current SFR in the galaxy is $\sim$ 2 M$_{\odot}$/yr.

\section{The H$\alpha$ tail}

The H$\alpha$ tail behind \ga\ is significant in the first 1$'$ from the
nucleus, even without continuum subtraction (Fig. 1).
Nevertheless, the high density of the foreground stars makes the
detection of faint, diffuse H$\alpha$ emission not easy in this region.
In our analysis, bright stars are masked and the rescaled H$\alpha_{\rm off}$ image
is subtracted. The net H$\alpha$ image shows a tail to at least 2.2$'$
(or 40 kpc) from the galaxy. We notice that the end of the
X-ray tail is close to the brightest star in the field, which makes
detection of faint, diffuse H$\alpha$ emission very difficult there.
The width of the H$\alpha$ tail ranges from 3 - 4 kpc.
The H$\alpha$ tail aligns with the X-ray tail very well and both of them
are brighter in the first 1$'$ from the nucleus (the ``Head'' region of
the X-ray tail discussed in S06) than in regions beyond.

The total H$\alpha$ flux in the tail is measured in an 106$''\times18''$ box
covered the whole tail. This box is adjacent to the 20$''\times26''$ aperture
used in $\S$4.2 to measure the total H$\alpha$ flux in \ga.
Emission-line objects and stars in the tail are masked and the H$\alpha$ fluxes
in their positions are recovered from their immediate surroundings.
Without correction for intrinsic extinction, the total H$\alpha$
flux in the tail is 4.4$\times10^{-14}$ ergs s$^{-1}$ cm$^{-2}$ and the
total luminosity is 2.4$\times10^{40}$ ergs s$^{-1}$ (with the same conversion
factor used for the H$\alpha$ emission of the galaxy in $\S$4.2). The H$\alpha$
surface brightness in the tail ranges from 7.4$\times10^{-17}$
ergs s$^{-1}$ cm$^{-2}$ arcsec$^{-2}$ near the galaxy to 1.6$\times10^{-17}$
ergs s$^{-1}$ cm$^{-2}$ arcsec$^{-2}$ in the faint region, with an average
of $\sim$ 2.8$\times10^{-17}$ ergs s$^{-1}$ cm$^{-2}$ arcsec$^{-2}$. This
average surface brightness is similar to that of D100's H$\alpha$ tail in Coma cluster
(0.5 - 4 $\times10^{-17}$ ergs s$^{-1}$ cm$^{-2}$ arcsec$^{-2}$, Yagi et al. 2007)
and 5 - 10 times higher than those of H$\alpha$ tails behind two irregular galaxies in A1367
(Gavazzi et al. 2001). In Coma and A1367, the surface density of foreground
stars is a lot smaller, which permits detection of faint H$\alpha$ features.
We follow Yagi et al. (2007) to estimate the mass of the H$\alpha$ tail.
Assuming a cylinder with a 3.5 kpc diameter and a 40 kpc length (note the tail
is somewhat twisted), the rms electron density in the tail is $\sim$
0.045 cm$^{-3}$. The total mass is then 5$\times10^{8}$ M$_{\odot}$ for
a filling factor of unity. This amount of mass for 10$^{4}$ K gas in the tail
is similar to the X-ray gas mass of the tail in the same portion, $\sim$
6$\times10^{8}$ M$_{\odot}$ from S06. However, both the 10$^{4}$ K gas and the
10$^{7}$ K gas in the tail can be very clumpy. If the filling factor for the
H$\alpha$ emitting gas is 0.05, the total mass of the H$\alpha$ tail
reduces to 10$^{8}$ M$_{\odot}$. Nevertheless, the stripped ISM in the tail
accounts for a significant portion of the original ISM in the galaxy. The tail
may also have cooler gas component. \ga\ was undetected in the HI observations with
ATCA by Vollmer et al. (2001). However, the limit, $\sim 10^{9}$ M$_{\odot}$,
is rather high, largely because of the nearby (14.5$'$ away) bright radio
galaxy PKS 1610-60.
Future more sensitive radio HI and infrared observations are required to
better quantify the ISM content in the galaxy and in the tail.

\section{Discussion}

\subsection{The formation of HII regions in the halo and the tail}

Our observations reveal at least 29 HII regions, all downstream of \ga.
Interestingly, the HII regions closest to the galactic disk form
a bow-like front with the axis close to the tail. Their projected distances
from the galactic nucleus are up to 39 kpc.
Most of them (if not all of them) appear away from the galactic disk plane.
Similar intracluster HII regions have been found before in the Virgo
cluster: one HII region 21 kpc from NGC~4388's nucleus in projection
(Gerhard et al. 2002) and one HII region 3 kpc off the disk and 6.5 kpc from
NGC~4402's nucleus in projection (Cortese et al. 2004).  
Both galaxies are 2 - 3.2 times more luminous than \ga\ in the $K_{s}$ band,
while their isolated HII regions have H$\alpha$ luminosities of only
1.3$\times10^{37}$ ergs s$^{-1}$ and 2.9 $\times10^{38}$ ergs s$^{-1}$
respectively (intrinsic extinction corrected). Intergalactic HII regions in
poorer environments have also been found (Ryan-Weber et al. 2004;
Mendes de Oliveira et al. 2004).
However, none of these known examples match the high number density and
luminosities of HII regions downstream of \ga. For example, 17 HII regions found
in this work have H$\alpha$ luminosities of $> 3\times10^{38}$ ergs s$^{-1}$
before correction for intrinsic extinction, while only one in previous
work (the brightest one in Mendes de Oliveira et al. 2004) is this luminous.
Moreover, five HII regions found in this work are up to five time more
luminous than the brightest one in Mendes de Oliveira et al. (2004).
Thus, the isolated star formation activity downstream of \ga\ is unprecedented.

As \ga\ traverses A3627, its dark matter halo is tidally truncated by the
cluster tidal field. As we don't know any characteristic
velocity of the galaxy, only a crude estimate can be made. 
In the simulations by Gnedin (2003), a dark matter halo of a large spiral galaxy
with a circular velocity of 250 km/s is truncated at 30$\pm$6 kpc in a cluster
similar to the Virgo cluster ($\sigma_{\rm radial}$ = 660 km/s). In smaller
galaxies, the truncation radius is proportional to the circular velocity.
From the Tully-Fisher relation (e.g., Gnedin et al. 2007), a disk galaxy with
\ga's stellar mass has a typical circular velocity of $\sim$ 100 km/s.
Thus, we take a conservative estimate of 15 kpc for the tidal truncation radius
of \ga's dark matter halo. This estimate is also consistent with the analytical
estimate by Merritt (1988). As shown in Fig. 1, most HII regions may still be
in the halo, although three objects at 29 - 39 kpc from the galaxy in
projection are hardly
still bound. Without velocity measurements, it is unknown whether most HII regions are
still bound in the galactic halo. However, they may easily escape \ga's
potential via weak tidal interaction that barely affects the galactic disk.
We also notice that both the H$\alpha$ and X-ray tails are brightest in the halo.

How did these HII regions form?
There is no indication in the optical that \ga\ is merging with
another galaxy. ELO1 is an emission-line source with insignificant
continuum (Fig. 1 and Table 1). In fact, the merger rate in a massive cluster
like A3627 should be very small as the cluster velocity dispersion is
high ($\sigma_{\rm radial}$ = 925 km/s for A3627).
Within 100 kpc (or 5.3$'$) from the end of the X-ray tail, there are
only two other galaxies identified from DSS2, 2MASS and our data,
WKK~6166 and another small galaxy without NED ID (G1 and G2 in Fig. 1b).
Neither has velocity
information in the literature, and both are in the FOV of our SOAR data.
In the $I$ band, WKK~6166 is 1.6 - 1.9 mag fainter than \ga, depending on
whether its center is blended with a star. Its radius is only $\sim 10''$,
or 3.2 kpc if it is in A3627. The other galaxy (G2 in Fig. 1b) is blended with a star.
In the $I$ band, it is 2.5 - 2.7 mag fainter than \ga, with a major axis
extending to $\sim 7.5''$. As both galaxies are redder than \ga\ with $B - I$
of 1.9 - 2.0 mag, their stellar mass is $\sim$ 30\% - 75\% of \ga's, from
the mass-to-light ratios in Bell \& De Jong (2001). However, even if these two
galaxies are in A3627 and have had recent fly-bys with \ga\ to tidally strip the gas
clouds responsible for these HII regions, stars should also be tidally stripped.
We would expect both stellar trails and the HII regions to be present in both directions
from \ga. On the contrary, the HII regions are present only in
a 120 deg cone downstream of \ga. There are some continuum features downstream of \ga\
up to 6 kpc from the nucleus
(Fig. 1 and Fig. 6). However, all of them have bright HII regions within or nearby.
The upstream side of the galaxy appears undisturbed and free of any tidal features
in both the $B$ and $I$ bands.

Cluster galaxies also undergo tidal interaction with the cluster potential
(e.g., Gnedin 2003).
Recently, Cortese et al. (2007) reported two peculiar galaxies in the massive
clusters A1689 ($z = 0.18$) and A2667 ($z = 0.23$). \hst\ images reveal
stellar trails composed of bright blue knots and fainter streams behind
both galaxies extending to 20 kpc and 75 kpc respectively. 
The fainter galaxy 131124-012040 in A1689 has a similar NIR luminosity to \ga,
although the star formation in the galaxy has stopped $\sim$ 100 Myr ago.
They argue that
these features are produced through tidal interaction with the cluster potential.
However for \ga, there are no stellar trails connecting the HII regions, at least
in the current data. The detected stellar features 6 kpc downstream of the
galaxy are confined around the emission-line regions and hardly extend much. Thus,
it is unclear how tidal interactions can strip the clouds responsible for
the current HII regions and star clusters, yet not produce stellar trails.
Future observations, like optical spectroscopy along both axes and \hst\ imaging,
will allow us to further quantify the dynamical state of \ga\ (e.g., how dynamically
cool is the disk?) and the stellar diffuse emission downstream of the galaxy.

The significance of ram pressure stripping in \ga\ and the bow-like front of the
HII region distribution drive us to consider another formation mechanism
for these HII regions, star-formation in the ram-pressure displaced ISM clouds.
\ga\ is undergoing a strong interaction with the surrounding ICM, judging
by the compactness
of the remnant H$\alpha$ disk. Assuming an ambient ICM density of 6$\times10^{-4}$
cm$^{-3}$ and an ICM temperature of 6 keV (see S06), and for a velocity of the galaxy
of 1600 km/s ($\sim \sqrt{3} \sigma_{\rm radial}$), the thermal pressure is
0.4$\times10^{5}$ K cm$^{-3}$, while the ram pressure is 1.1$\times10^{5}$ K cm$^{-3}$.
Therefore, the ram pressure is high enough to strip 10 K - 100 K ISM clouds with
density up to 10$^{4}$ cm$^{-3}$. HI gas will be stripped, as well as some less
dense dark clouds and the outskirts of molecular clouds, even though the dense cores of
molecular clouds may remain in the disk plane.
Schaye (2004) found a critical surface density of $N_{H} = (3-10) \times 10^{20}$
cm$^{-2}$ for star formation in the outer parts of the galactic disks.
Some stripped dense clouds from the disk of \ga\ should have higher
surface density than this critical value, even if clouds are very clumpy.
Thus, the ability of high ram pressure to displace dense clouds from the disk
may affect the efficiency of star formation in the stripped ISM.
The initial instantaneous stripping can be fast in a high ram pressure
environment ($\gsim$ 10 Myr from RH05), while most materials stripped
off the disk still remain bound. The subsequent evolution of the stripped
ISM is complicated. Simulations (Vollmer et al. 2001; SS01; RH05) show that a fraction of
stripped materials (mainly from the outer disk, $\sim$ 10\% from RH05) can
still be bound in the galactic halo (the ``hang-up'' effect) for a few
100 Myr even in a high ram pressure
environment. Some material may even fall back to the disk, while most of the
stripped ISM concentrates within several galaxy diameters downstream, in
the lee of the galaxy where the ram pressure is reduced (e.g., SS01).
The details of the process rely on many factors, for example, the porosity of
the ISM in the inner disk or the effective ram pressure downstream of the
galaxy, the halo potential, and the efficiency of heat conduction and viscosity.
Simulations (RH05; Roediger, Br\"{u}ggen \& Hoeft 2006) also show that
the wake can be very turbulent. The stripped clouds will collide with
each other, which may produce shocks and
trigger star formation. Moreover, being in the halo separates them from the
strong stellar UV heating flux present in the disk, allowing cooling
to be dominant. Thus, star formation may proceed in the halo or even in
unbound clouds far away from the galaxy (e.g., ELO8 and 9).
We notice that Oosterloo et al. (2004) reported that two intergalactic HII regions
at least 100 kpc from the nearest galaxy (an elliptical) are in a massive
isolated HI cloud. HII regions in HI clouds were also reported in Stephan's
quintet by Mendes de Oliveira et al. (2004).
Thus, star formation in isolated HI clouds is plausible, although the
efficiency may be low as suggested by Oosterloo \& van Gorkom (2005).
This scenario can explain the spatial distribution of the HII regions
although the details are certainly complicated.
SS01 even suggested formation of dwarf galaxies by these means.
We indeed notice that the brightest HII region, ELO1, has an estimated total
mass of $\sim 4\times 10^{6}$ M$_{\odot}$. If star formation in ELO1
is still active, ELO1 may grow into a dwarf
galaxy eventually.
Future \hst\ and high-resolution HI observations are required to better
understand ELO1, other bright HII regions and the stripped ISM.

\subsection{The fate of the isolated HII regions and their implications}

The estimated ages of the HII regions are less than $\sim$ 8 Myr.
The length of the X-ray tail implies that stripping has lasted for $\sim$
50 Myr (assuming a velocity of $\sim$ 1500 km/s in the plane of the sky).
Thus, it is possible that many HII regions may have already faded.
Nevertheless, the long spatial distribution of the HII regions along
the tail implies that after the ISM has been stripped, the time lapse
to the beginning of star formation spans a range up to $\sim$ 20 Myr or more
(e.g., for ELO8 and 9). Contrary to ELO8 and 9,
some luminous HII regions (e.g., ELO1 - ELO3) are only $\sim$ 1.6 - 5 kpc
from the galactic disk in projection, but their ages are comparable to
those of ELO8 and 9 (if not longer). Their estimated ages are longer than
the required time to produce the observed projected offsets. Thus, this
may be the evidence that a significant part of stripped ISM from the
disk can still be bound in the halo.

We also explore the possibility of detecting faded HII regions from the
imaging data. The six additional emission-line regions selected
with looser criteria (Fig. 2) are certainly candidates with stronger
$I$ band continuum. They are also close to the 29 HII regions.
The embedded star clusters are blue if the intrinsic extinction is not too
high. From Starburst99 model, we find that star clusters like those in ELO1 - 3
should be detected in $B$ and $I$ bands with our data, even at ages of
50 - 60 Myr. However, it is difficult to distinguish them from the various types
of Galactic stars in the foreground. About 10 Myr after the initial starburst,
red supergiants appear and the galactic colors are highly metallicity dependent
over this period. For star clusters with metallicities larger than 0.4 solar,
Starburst99 also predicts a big increase for the $B - I$ color at $\sim$
10 Myr for instantaneous starbursts. $B - I$ color can be as high as 1.7 mag
in that period, which makes it almost impossible to select them from Galactic
stars with imaging data alone (Fig. 2). In fact, ten non-emission-line objects
with $B - I <$ 0.8 mag (Fig. 2) are mostly bright stars and none of them are
close to the tail. Thus, faded HII regions may be selected only through a
spectroscopic survey.

If formation of massive stars proceeds in the stripped ISM, high-mass X-ray
binaries may form in those star clusters. The existing shallow (14 ks) \chandra\
observation reveals only three hard X-ray point sources in the tail ($\gsim 10^{40}$
ergs s$^{-1}$ if in A3627), but both the \xmm\ and the \chandra\ spectra reveal
an unresolved hard component in the tail (S06). Interestingly, the brightest
\chandra\ point source P1 (S06) is only $\sim 0.4''$ from the putative HII
region ELO5 (Fig. 1), while the combined positional uncertainty from \chandra\ and
the SOAR data can be up to $\sim 0.3''$. The \chandra\ point source P3 positionally coincides
with a red optical source ($B-I$=3.2) with an H$\alpha$ - H$\alpha_{\rm off}$
color of 0.01, which is likely a star. The \chandra\ point source P2 is
0.7$'' - 0.9''$ from two star-like objects, but only 2.5$''$ away from an
emission-line object candidate (Fig. 1). We have a 150 ksec \chandra\ ACIS-S
observation approved for cycle 9, which can probe the X-ray point source
distribution 20 times deeper than the existed data.

The \ga\ data strongly imply a connection between stripping of the ISM and 
star formation in the halo and intracluster space. The stripped
ISM not only contributes to the ICM, but also adds intracluster
light by subsequent star formation after stripping. The total mass of starbursts
in the 29 HII regions is about 10$^{7}$ M$_{\odot}$ (with large uncertainty from
age and extinction), while the total gas mass of the H$\alpha$ tail and
the X-ray tail is $\lsim 10^{9}$ M$_{\odot}$ (depending on
the filling factor and density variation). As many HII regions may have
already faded and star formation may still proceed in the stripped ISM,
the final total stellar mass formed in the halo and intracluster space
should be at least several times larger than the 10$^{7}$ M$_{\odot}$
estimated above.
As most of HII regions may still be in the halo,
it is possible that star formation in stripped ISM is more efficient in
the halo than in intracluster space.
For the star clusters in the halo, some of them may remain bound, while others
may escape the galactic potential through tidal interactions.
Thus, this single transforming galaxy \ga\ contributes over 10$^{7}$ M$_{\odot}$
of new stars in the halo and intracluster space through star formation in the
ram-pressure displaced clouds, although it is unclear how many will end
in intracluster space.
If some large star clusters are bound to the galaxy for a significant amount of
time, the configuration may look like galaxy aggregates found in the Coma cluster
(Conselice \& Gallagher 1998).

\subsection{H$\alpha$ tail and stripping of the galactic ISM}

There are not many cluster galaxies with known H$\alpha$ tails,
two irregular galaxies in the northwest of A1367 (Gavazzi et al. 2001),
NGC~4388 in the Virgo cluster (Yoshida et al. 2002; 2004), a
post-starburst galaxy near the center of the Coma cluster (Yagi et al.
2007), and several galaxies in an infalling compact group towards
A1367 (Sakai et al. 2002; Gavazzi et al. 2003; Cortese et al. 2006).
Both galaxies in the northwest of A1367 and the post-starburst galaxy
in Coma have similar $K_{s}$-band absolute
magnitudes as \ga's (from 0.49 mag fainter to 0.10 mag brighter), implying
similar total stellar mass as \ga's. NGC~4388 is 1.3 mag more luminous than
\ga\ in the $K_{s}$ band and also larger. Both H$\alpha$ tails in A1367 are
suggested to be produced by ram pressure stripping, while the tidal
interaction between the two galaxies may trigger star formation in both
galaxies (Gavazzi et al. 2001). The H$\alpha$ tail behind the galaxy D100
in Coma is remarkably narrow and straight (60 kpc $\times 2$ kpc).
It may be produced through either ram pressure stripping, or comes from
the gas of a merging dwarf (Yagi et al. 2007). The H$\alpha$ filaments in
the northeast of NGC~4388 are suggested to be ionized by NGC~4388's AGN,
while ram pressure stripping of the ISM provides the cool gas trail
(Yoshida et al. 2004). The remarkable infalling compact group in A1367 is
composed of several giant galaxies, many dwarf galaxies and H$\alpha$-emitting
knots. We notice that the $B$ band absolute magnitude of the ELO1 (before
correction on the intrinsic extinction) is comparable to
the faint dwarf galaxies in that infalling group (M$_{B}$ = -15.3 mag compared
to -15.8 mag). Cortese et al. (2006) suggested that the rich star formation
phenomena in this infalling group is triggered by tidal interactions between
group members and the ram pressure by the ICM.
The galaxy C153 in A2125 ($z$=0.247) is another similar case (Wang et al. 2004;
Owen et al. 2006), with a wide 80 kpc [OII] tail that is also likely shown
in the X-rays. The galaxy has a distorted disk with substantial star formation
activity, just when it is penetrating the cluster core. However, it is
far more distant than all other galaxies discussed here so the interaction
cannot be studied in the same detail as others.
There are no detections of X-ray tails behind the other
galaxies with H$\alpha$ tails in the \chandra\ or \xmm\ data.
There are also no reports of isolated HII regions near
the two galaxies in the northwest of A1367 and D100 in Coma. The
spectroscopically confirmed HII region by Gerhard et al. (2002) is
north of NGC~4388 but is far away from the H$\alpha$ filaments revealed
by Yoshida et al. (2002). Other emission-line objects in that part of
the Virgo cluster are even farther away from the H$\alpha$ filaments
(Fig. 1 of Okamura et al. 2002).

How does the H$\alpha$ tail of \ga\ form? 
The net H$\alpha$ image of the galaxy shows the truncation of the H$\alpha$
emission on the front and the sides (Fig. 1 and Fig. 6). In fact, the
leading edge of the X-ray emission in \ga\ coincides positionally with
the H$\alpha$ front. 
The compactness of the remaining H$\alpha$ core demonstrates the
strength of the ICM wind. Compared with simulations by Roediger \&
Hensler (2005), over 90\% of the gas in the original gas disk should have
been removed from the disk, either hanging up in the halo or in the
tail. Thus, ram pressure stripping should play an important role
in the formation of the observed H$\alpha$ tail.
The stellar ionization field in the tail may be too weak to ionize the stripped
cold gas to produce the observed H$\alpha$ tail, at least from the current data.
The positional coincidence of the H$\alpha$ tail and the X-ray tail implies
a connection.
Can heat conduction from the surrounding ICM to the cold stripped ISM
provide enough energy?
Heat conduction across the embedded cold clouds is most likely saturated.
We use the saturated heat conductivity derived by Cowie \& McKee (1977)
to estimate the flow-in heat flux. The surface area is still estimated
from a cylinder with a 3.5 kpc diameter and a 40 kpc length ($\S$5), although
the real contact surface is almost certainly higher (e.g., for clumpy
clouds). The total heat luminosity is 4.7$\times10^{43}$ ergs s$^{-1}$,
for an ICM temperature of 6.5 keV and an ICM density of 10$^{-3}$ cm$^{3}$.
Although the energy emitted through H$\alpha$ is only a small fraction of the total
optical line cooling (several percent from Voit, Donahue \& Slavin 1994),
there is enough heat energy for the observed H$\alpha$ tail.
However, heat conduction can be largely suppressed by magnetic field
at the boundaries of the clouds. It is likely that both stripping and
heat conduction contribute to the observed H$\alpha$ tail.
Spectroscopic studies of the H$\alpha$ tail, though difficult, will better
clarify its nature.
Radio observations are also required to study the distribution of the cold
atomic and molecular gas in and behind \ga.

\section{Conclusions}

\ga\ is a star-forming galaxy with a dramatic X-ray tail (S06).
In this work, we present optical imaging and spectroscopic data of
\ga\ from SOAR and the CTIO 1.5m telescopes.
An H$\alpha$ tail is found behind \ga, extending to at least
40 kpc from the galaxy. The H$\alpha$ tail positionally coincides
with the X-ray tail. We conclude that ram pressure stripping is
responsible for the ISM tail, while heat conduction from the hot ICM
may also contribute to the energy of the optical lines. This discovery makes \ga\ the only
known cluster galaxy with an X-ray tail and an H$\alpha$ tail.
The H$\alpha$ emission in \ga\ is confined to the central $\sim$ 1 kpc
(radius) region. The current nuclear SFR ($\sim$ 1.2 M$_{\odot}$/yr) and
the largely enhanced stellar continuum (after correction of intrinsic extinction)
around the center imply that a galactic bulge may be building up.
At least 29 emission-line objects with high H$\alpha$ flux
are also revealed by the data. They all spatially cluster
downstream of the galaxy, in or around the stripped ISM tail.
Those closest to the galactic disk form a bow-like front. From
analysis of the H$\alpha_{\rm off}$ frame and the estimate of the
background emission-line objects, we conclude that the expected number
of background objects is much less than one in the whole
3.43$'\times3.88'$ field and all 29 emission-line objects are very likely
HII regions associated with \ga. The high number density and luminosities
of these HII regions downstream of \ga\ dwarf the previously
known examples of isolated HII regions in clusters, making \ga\
a dramatic example full of rich phenomena. Interestingly, one \chandra\
point source (an ULX with $L_{X} > 10^{40}$ ergs s$^{-1}$ if in A3627)
is in the same position as an HII region.
All these detections indicate significant star formation activity
in the halo of \ga\ and in intracluster space.

Our data imply a connection between these HII regions and
the stripping of the ISM. We suggest that star formation may proceed
in the stripped ISM, in both the galactic halo and the intracluster
space. The star formation in the halo may have a higher efficiency as
the stripped ISM is still bound. Cooling can become dominant when the
surrounding UV radiation field is much weaker than in the disk. Cloud
collisions may also induce collapse of the ISM clouds. The total mass
of the current starbursts in these 29 HII regions is about 10$^{7}$
M$_{\odot}$. Since many HII regions and star clusters may have already
faded, the total stellar mass formed in the stripped
ISM may be several times higher. Therefore, stripping of the ISM
not only contributes to the ICM, but also adds intracluster
stellar light through subsequent star formation.
 
There is no reason to believe \ga\ is unique. However,
its current active stage may last only for a short time ($< 10^{8}$ yr)
so a large sample of cluster late-type galaxies has to be studied to
find more similar ones like \ga, either in X-rays or H$\alpha$.
Nevertheless, the proximity of \ga\ makes it a good target to
study galaxy transformation, the evolution of stripped ISM,
star formation and X-ray binaries in the halo and the intracluster space,
with the help of more data.

\acknowledgments

M. S. is very grateful to Allan Hornstrup, who made an early H$\alpha$
observation using the Danish 1.54m telescope with observers Jan-Erik Ovaldsen,
Josefine H. Selj, and Dong Xu.
We thank Nathan De Lee for the help on the SOAR data analysis.
We thank Fred Walter for the helps
on the CTIO 1.5m data and the observation planning.
We acknowledge the work of the SOAR operators Sergio Pizarro and Daniel
Maturana, as well as the CTIO 1.5m service observers Alberto Miranda and Alberto
Pasten. We thank the referee, Giuseppe Gavazzi, for helpful and prompt comments.
The financial support for this work was
provided by the NASA Grant AR6-7004X and NASA LTSA grant NNG-05GD82G.
We are presenting data obtained with the Southern Observatory for
Astrophysical Research, which is a joint project of the Brazilian
Ministry of Science, the National Optical Observatories, the University
of North Carolina and Michigan State University. We made use of the
NASA/IPAC Extragalactic Database and the Hyperleda database.

\clearpage

\begin{table}
\begin{small}
\begin{center}
\caption{Properties of representative emission-line objects (putative HII regions)}
\vspace{-1mm}

\begin{tabular}{ccccccccc}
\hline \hline  ID\tablenotemark{a} & $L_{H\alpha}$\tablenotemark{b} & $Q(H^{0})$\tablenotemark{c} & $M_{B}$\tablenotemark{d} & $M_{I}$\tablenotemark{d} & EW (H$\alpha$)\tablenotemark{e} & Age\tablenotemark{f} & Mass\tablenotemark{g} & N(O)\tablenotemark{h} \\
 & (10$^{39}$ ergs s$^{-1}$) & (10$^{51}$ s$^{-1}$) & (mag) & (mag) & (\AA) & (Myr) & (10$^{5}$ M$_{\odot}$) & (10$^{2}$) \\
\hline

1 & 21 & 21 & -15.29 & -16.76 & 138 & 3.4, 5.4, 6.0, 6.0 & 38 & 89 \\
2 & 8.6 & 8.5 & -13.62 & -13.82 & 614 & 2.7, 3.2, 4.7, 4.7 & 5.3 & 20 \\
3 & 6.8 & 6.6 & -13.32 & -13.92 & 437 & 2.6, 3.5, 4.8, 4.8 & 5.9 & 21 \\
4 & 1.4 & 1.4 & -13.39 & -14.17 &  68 & 5.0, 5.7, 6.7, 6.5 & 4.6 & 7.7 \\
5 & 1.0 & 0.99 & -12.00 & -12.42 & 281 & 3.4, 4.7, 5.1, 5.0 & 0.95 & 3.4 \\
6 & 0.096 & 0.095 & $>$-10.48 & $>$-11.02 & $>$90 & $<$5.0, $<$5.4, $<$6.3, 3.0 & 0.019 & 0.07 \\
7 & 1.3 & 1.3 & -11.28 & $>$-11.94 & $>$320 & 2.3, $<$3.4, $<$5.0, 3.5 & 0.43 & 1.9 \\
8 & 0.10 & 0.10 & $>$-10.59 & $>$-11.14 & $>$95 & $<$5.0, $<$5.5, $<$6.3, 5.0 & 0.096 & 0.34 \\
\hline \hline

\end{tabular}
\tablenotetext{a}{Source ID (see Fig. 1c and 1d)}
\tablenotetext{b}{1 mag intrinsic extinction assumed (see $\S$3.3 for more detail)}
\tablenotetext{c}{The number of H-ionizing photons estimated from $L_{H\alpha}$}
\tablenotetext{d}{Absolute magnitudes without correction on intrinsic extinction}
\tablenotetext{e}{The equivalent width of H$\alpha$ line assuming an [NII] fraction of 19\% (see $\S$3.3)}
\tablenotetext{f}{The age of the current starburst estimated from Starburst99 (see $\S$3.3).
The first three values are estimated from $Q(H^{0})/L_{B}$, $Q(H^{0})/L_{I}$ and EW (H$\alpha$),
while the final value is the chosen age to estimate the next two properties. We generally use
the age derived from EW (H$\alpha$) if it is well determined, as EW (H$\alpha$) is not
affected by intrinsic extinction. Different ages for similar objects \#6 and \#8 are chosen
to demonstrate the change of the total mass and the number of O stars with the choice of the
starburst age.}
\tablenotetext{g}{The total mass in the instantaneous starburst estimated from Starburst99}
\tablenotetext{h}{The number of O stars estimated from Starburst99}
\end{center}
\end{small}
\end{table}

\clearpage
\pagestyle{empty}
{\epsscale{1}\plotone{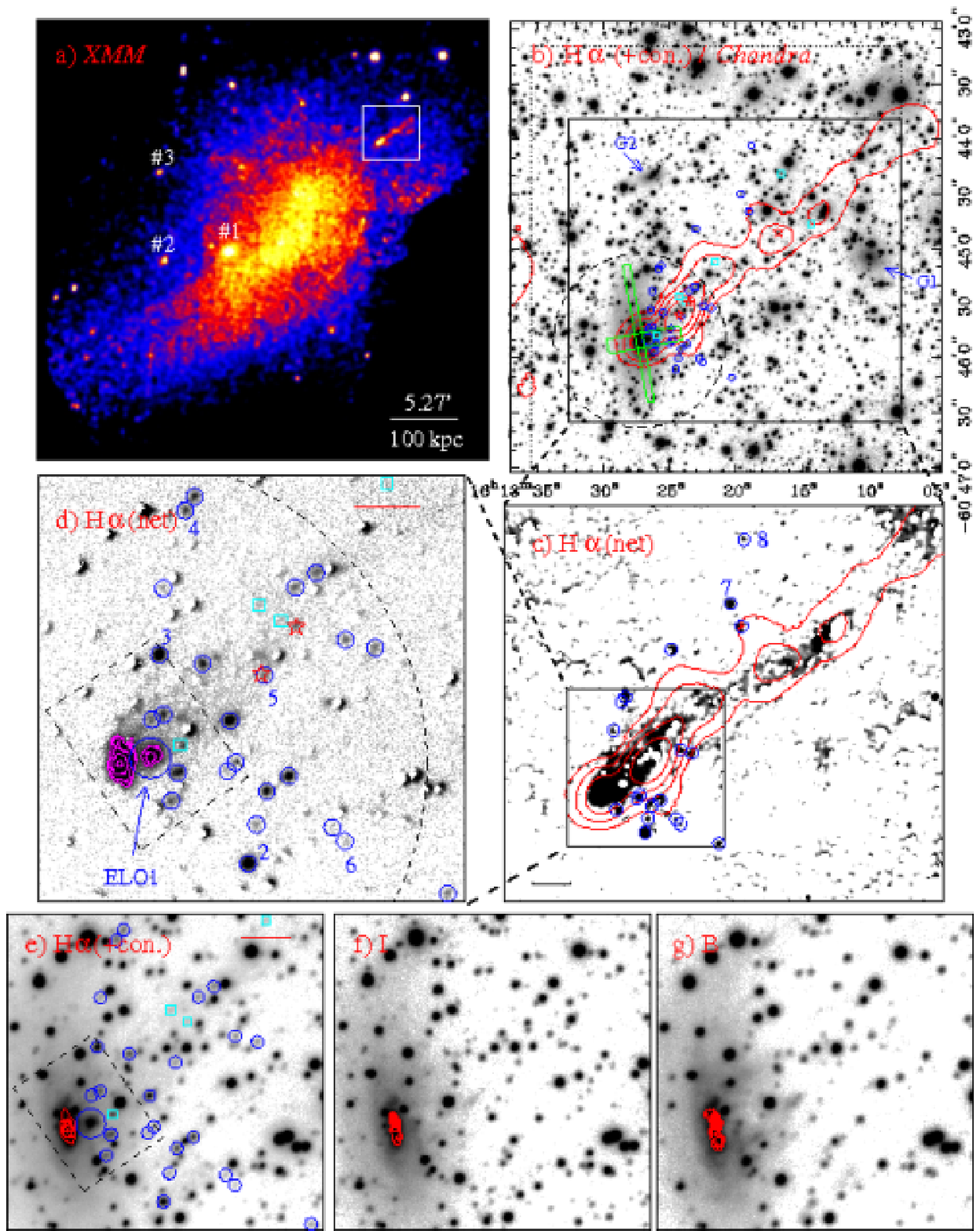}}
\clearpage
\begin{figure}
\caption{
{\bf a)}: \xmm\ 0.5 - 2 keV mosaic of A3627 (background-subtracted and
exposure corrected). A long X-ray tail is significant as shown in the
white box. The brightest three cluster galaxies (in the $K_{s}$ band) are
away from the cluster gas core
(ESO~137-006 - \#1, ESO~137-008 - \#2 and ESO~137-010 - \#3).
{\bf b)}: the zoom-in of the region around \ga's tail (4.0$'\times4.1'$).
\chandra\ contours (from the 0.5 - 2 keV smoothed image, in red) are
superposed on the H$\alpha$ (+ continuum) image. The dotted box region is
where data in H$\alpha$, H$\alpha_{\rm off}, B$ and $I$ bands are all available and
colors are measured in Fig. 2. Three \chandra\ point sources in the tail
are also marked as red stars. Blue circles mark emission-line objects
selected from colors in Fig. 2, while cyan boxes mark six more sources
selected if criteria to select emission-line objects are relaxed ($\S$3.1). The green
rectangles mark the regions where surface brightness profiles are measured
along the minor and major axes shown in Fig. 6.
The dashed-line circle (15 kpc in radius) shows the approximate size of the tidally
truncated dark matter halo of \ga\ (also in d).
Most of the emission-line objects may still be in the halo, but the most distant three are
29 - 39 kpc from the nucleus. G1 and G2 are the only two galaxies within 100 kpc
from the end of \ga's X-ray tail (velocity unknown). The cluster is close to the Galactic plane
(Gal. latitude of -7 deg) so the foreground star density is high. 
{\bf c)}: net H$\alpha$ emission (smoothed) in a contrast to enhance the emission
from the diffuse tail, superposed on the \chandra\ contours (red).
Stars are masked. Beyond the H$\alpha$ tail and the emission-line objects (in
blue circles for those outside the tail), the residual emission is generally
around bright stars. The scale bar is 5 kpc (or 15.8$''$).
The X-ray leading edge in the current short \chandra\ exposure is actually
around the same position as the H$\alpha$ edge,
as the shown \chandra\ contours are from a smoothed image.
{\bf d)}: net H$\alpha$ emission in the galaxy and the head of the tail
(1.1$'\times1.1'$). The contours (in magenta) around the peaks are also shown.
Symbols have the same meaning as in b). Eight representative emission-line objects
are marked from ELO1 to 8 (sources 7 and 8 in panel c). The box
is the aperture used to measure the total net H$\alpha$ flux from the galaxy ($\S$4).
The scale bar is 10$''$ (or 3.17 kpc).
{\bf e)}: H$\alpha$ (+ continuum) image of the same region as d).
The nuclear region is also shown in red
contours (similar in the next two panels). The scale bar is 10$''$.
{\bf f)}: I band image in the same field as d).
{\bf g)}: B band image in the same field as d). Two blue features (from the
$B - I$ image) downstream of the galaxy are shown by arrows. There are also
broad-band features to the south of ELO1.
}
\end{figure}
\clearpage

\begin{figure}
\vspace{-1cm}
\hspace{1cm}
\includegraphics[scale=0.7,angle=270]{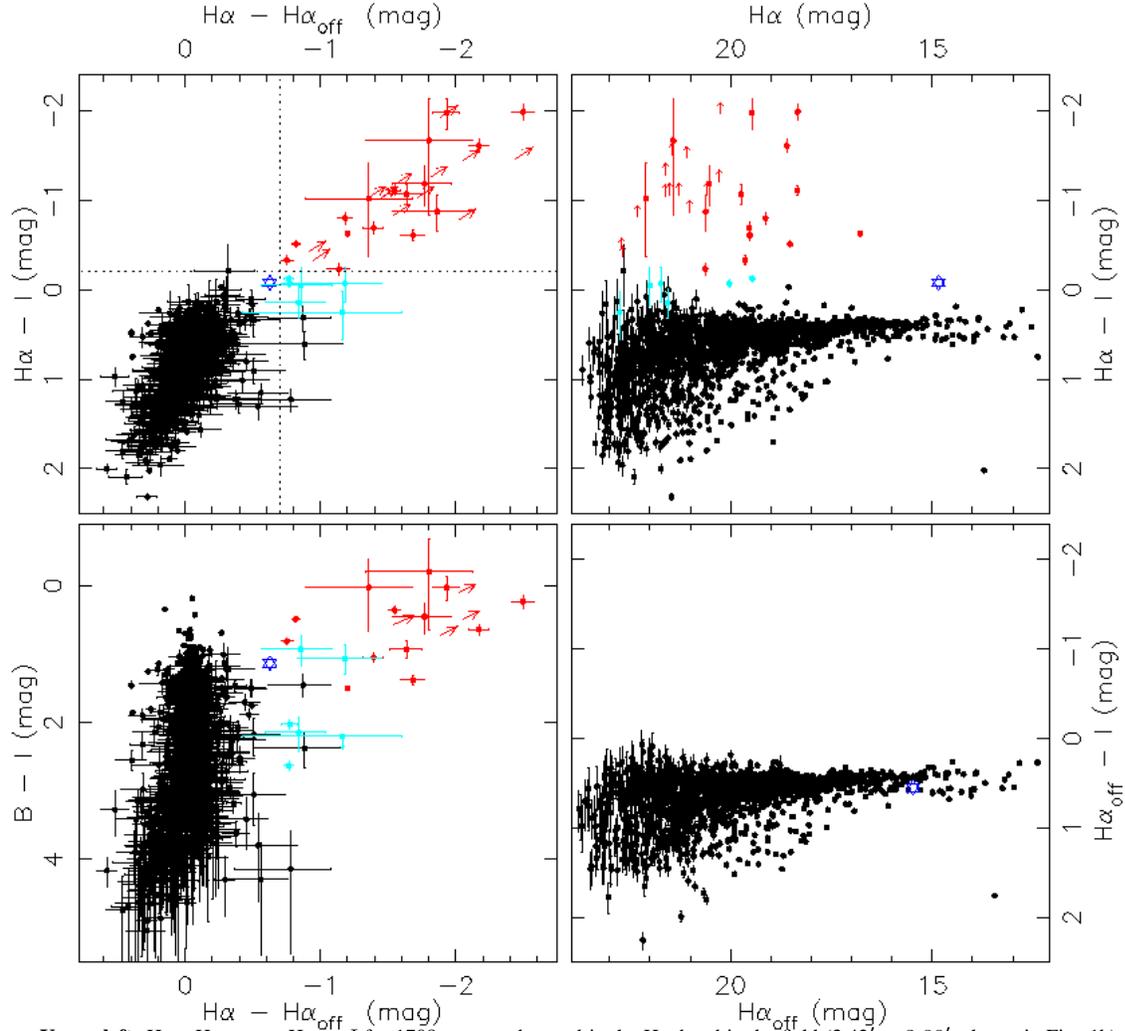}
\vspace{-0.1in}
\caption{
{\bf Upper left}: H$\alpha$ - H$\alpha_{\rm off}$ vs. H$\alpha - I$ for 1708 sources
detected in the H$\alpha$ band in the field (3.42$'\times3.88'$, shown in Fig. 1b).
Both the H$\alpha$ and the H$\alpha_{\rm off}$ fluxes include continuum. Over 97\% of
them are Galactic stars with H$\alpha$ - H$\alpha_{\rm off}$ color close to zero.
We define 29 emission-line objects (in red) as sources with
H$\alpha - I <$ -0.2 mag and H$\alpha$ - H$\alpha_{\rm off} <$ -0.7 mag (dotted lines).
The H$\alpha$ and H$\alpha_{\rm off}$ magnitudes are AB magnitudes (Hamuy et al. 1994).
We have corrected the Galactic extinction
for all sources (including Galactic stars, 0.15 mag reddening for the H$\alpha - I$
color) but no correction on the intrinsic extinction associated with \ga\ has
been applied (also the case for the next three plots).
The blue star marks the position of \ga's central region where the giant H$\alpha$
nebula is, measured within a 3$''\times6''$ (semi-axes) ellipse centered on the H$\alpha$
peak (also in the next three plots). If we relax the selection criteria (see $\S$3.1),
six more candidate emission-line objects are selected (in cyan).
{\bf Upper right}: H$\alpha$ magnitude vs. H$\alpha - I$ color.
{\bf Lower left}: H$\alpha$ - H$\alpha_{\rm off}$ color vs. $B - I$ color for sources
where the $B - I$ color can be constrained. The emission-line objects appear blue and most
of them are bluer than \ga's core (no correction on the intrinsic extinction).
The Galactic reddening for the $B - I$ color (0.49 mag) has been applied to all
sources.
With a small amount of intrinsic extinction, the $B - I$ colors of emission-line
objects are close to that for pure O/B stars (-0.2 to -0.8 mag).
{\bf Lower right}: H$\alpha_{\rm off}$ magnitude vs. H$\alpha_{\rm off} - I$ color for
sources detected in the H$\alpha_{\rm off}$ band. There are no emission-line
objects detected in the H$\alpha_{\rm off}$ band with H$\alpha_{\rm off} - I < -0.2$ mag. 
Moreover, no objects are detected with H$\alpha_{\rm off}$ - H$\alpha < $ -0.5 mag
and H$\alpha_{\rm off} - I < 0.25$ mag.}
\end{figure}

\clearpage
\pagestyle{plaintop}
\begin{figure}
\vspace{-0.5cm}
\hspace{0.2cm}
\includegraphics[scale=0.85,angle=270]{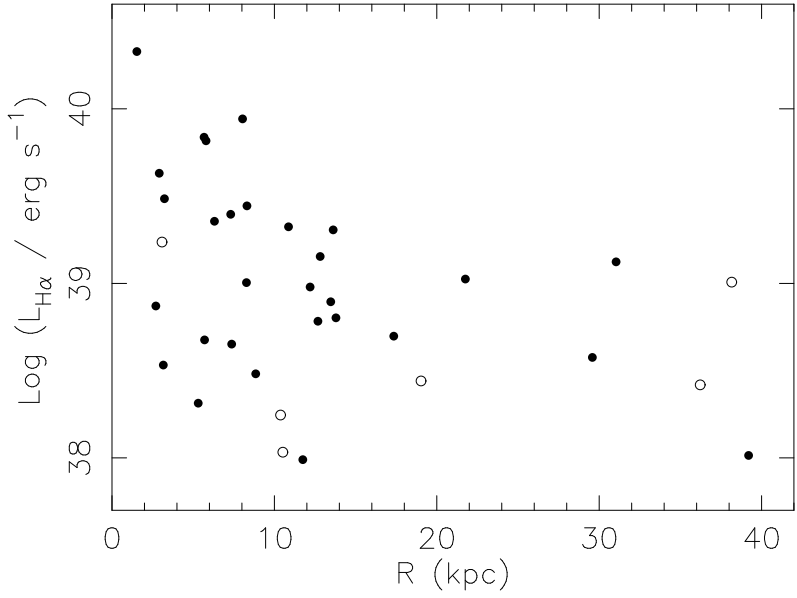}
\caption{The luminosities of HII regions plotted with their
distances to \ga's nucleus. The data points in empty circles are six
additional sources if the criteria for emission-line objects are relaxed ($\S$4).
We assumed 1 mag of intrinsic extinction for all of them.
For comparison, the two isolated HII regions in the Virgo
cluster (Gerhard et al. 2002; Cortese et al. 2004) have H$\alpha$
luminosities of 1.3$\times10^{37}$ ergs s$^{-1}$ and 2.9$\times10^{38}$
ergs s$^{-1}$ respectively (intrinsic extinction corrected).
}
\end{figure}

\begin{figure}
\vspace{-2cm}
\hspace{0.2cm}
\includegraphics[scale=0.86]{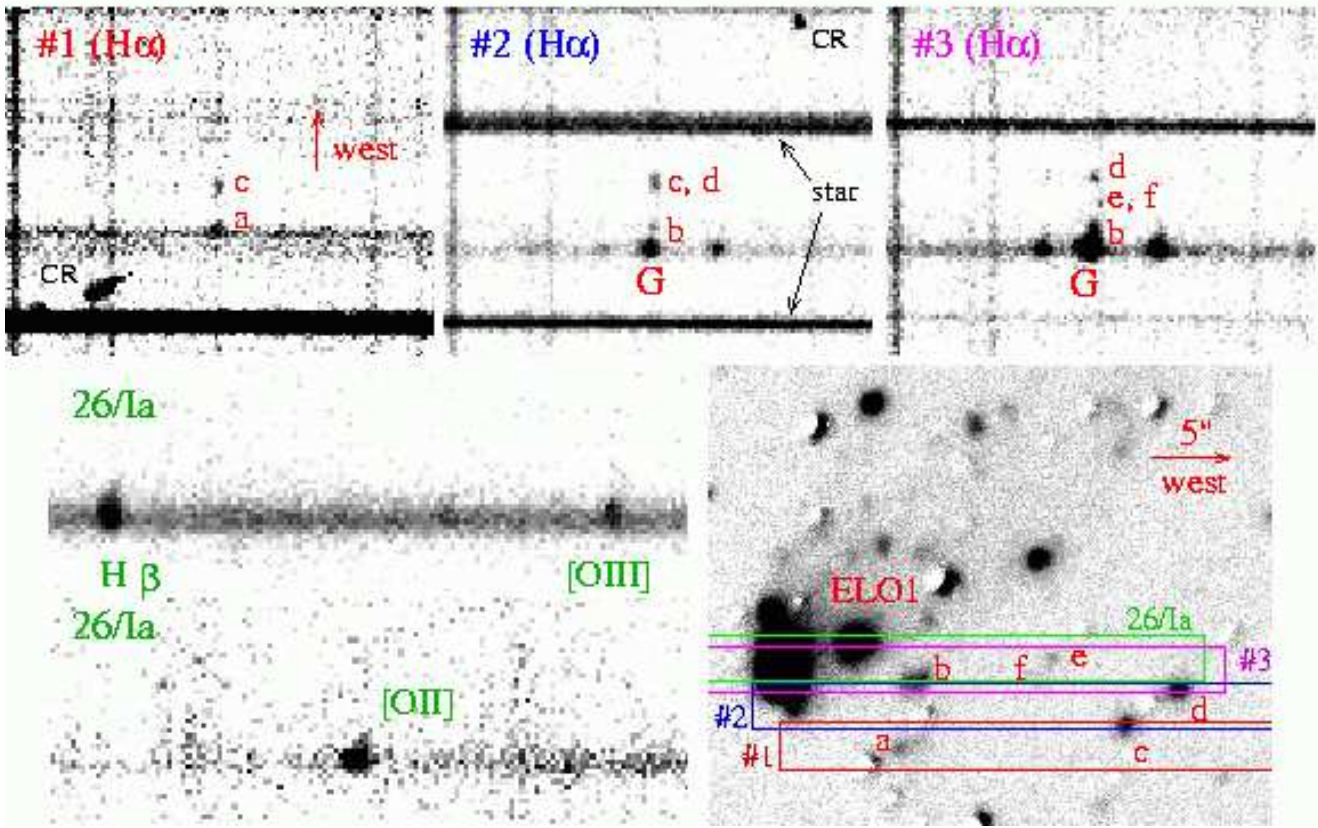}
\caption{The long-slit spectra from the three 47/Ib slit positions
and the 26/Ia slit position (in two wavelength ranges) shown in the upper panels and the lower left panel,
while these four slit positions are shown on the net H$\alpha$ image (the lower right
panel). The dispersion axis is always horizontal and the spatial axis is always
perpendicular on the detector plane (the west is upward).
Note that each slit has a length of $\sim 5'$ and a width of 3$''$
centered on the galaxy, while the slits plotted are truncated only for better
viewing. The three 47/Ib 2D spectra are spatially aligned. The \#1 position
completely misses the H$\alpha$ emission of the galaxy, while the \#3 position
covers the H$\alpha$ peak of the galaxy. Emission-line objects (ELO1, and a - f
as marked) were caught in one or more slit positions as shown in the spectra.
The small scale-bar in the \#3 plot (near ``G'') is 6.56\AA, or 300 km/s.
The slit for the 26/Ia observation
runs across the peak of the galaxy and a large part of ELO1. The 26/Ia spectrum
shows extension to the direction of ELO1 (upward on the detector plane) for the
three strong lines, which implies that
ELO1 has nearly the same velocity as \ga's. Note that the best instrumental
spatial seeing for the spectrograph is 3.5$''$ (or 2.7 pixels).
}
\end{figure}
\clearpage

\begin{figure}
\vspace{-1cm}
\hspace{0.2cm}
\includegraphics[scale=0.85]{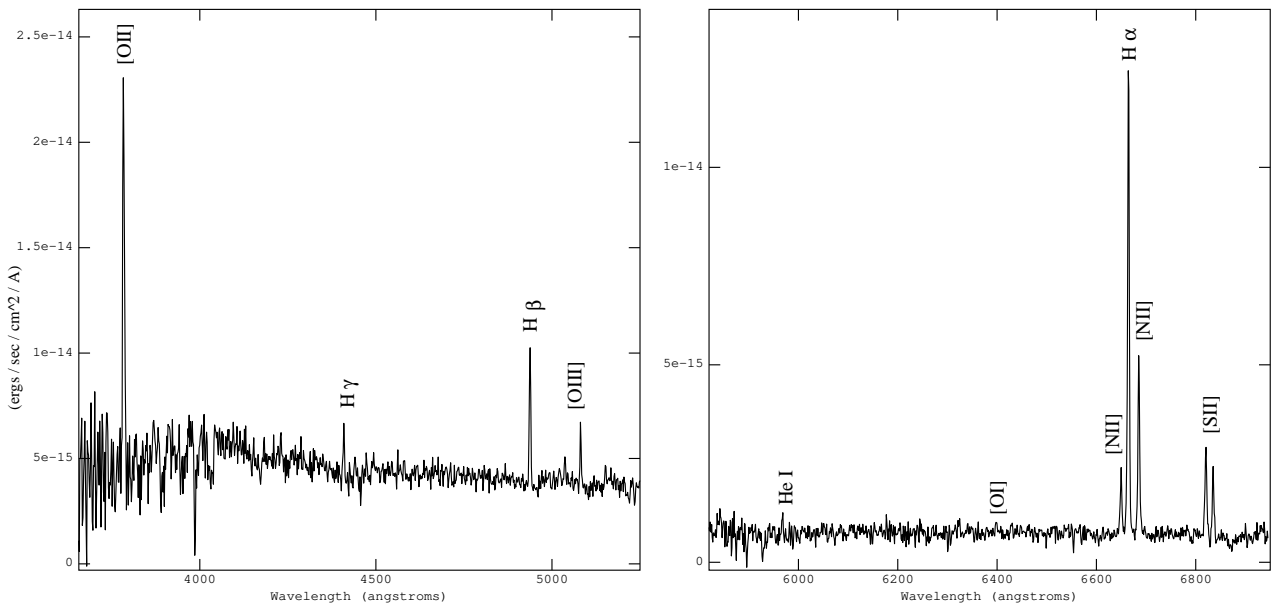}
\caption{The CTIO 1.5m spectra of \ga's core in two wavelength ranges.
The H$\alpha$ spectrum
was taken at the slit position \#3 (Fig. 4). Emission lines are marked.
The line widths are all small (e.g., 3.7 \AA\ FWHM for the H$\alpha$ line).
As the spectroscopic standards were observed with 2$''$ slits, the shown flux
of each spectrum is over-estimated by an unknown constant, but the spectral shape
and line ratios are not affected.
}
\end{figure}

\begin{figure}
\vspace{-1.5cm}
\includegraphics[scale=0.69,angle=270]{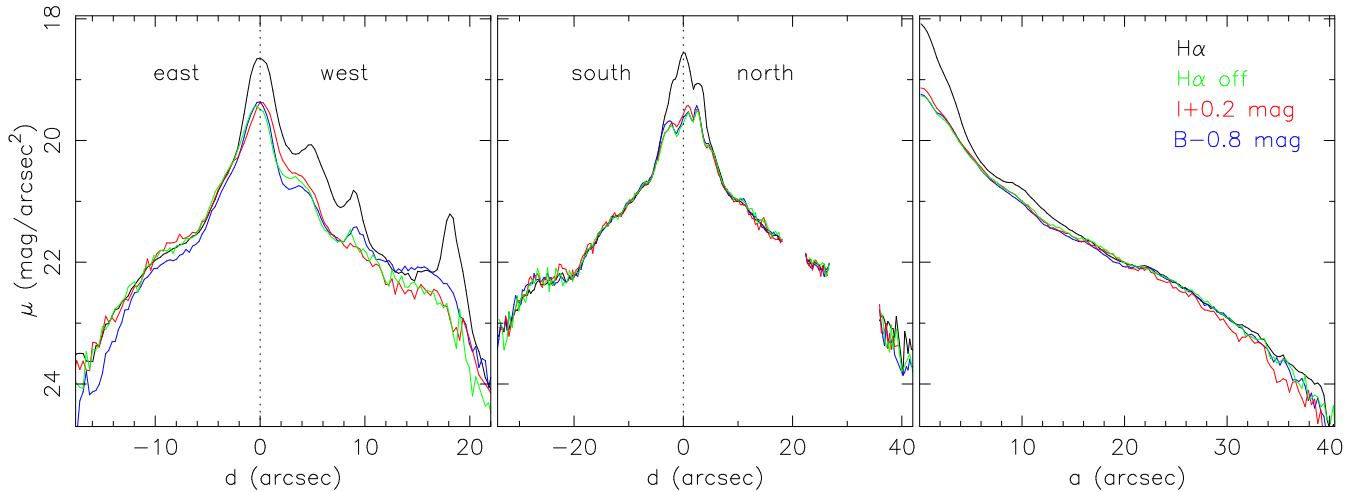}
\caption{{\bf Left}:
Surface brightness profiles measured parallel to the minor axis of the
galaxy (shown as a green box in Fig. 1b, 8$''$ width) in four bands
(color coding in the right plot).
We chose the origin as the peak of the H$\alpha$ profile (same in the next
two plots). Stars are masked but emission-line objects are
included (also for the plot shown in the middle panel). The H$\alpha_{\rm off}$
light traces the H$\alpha$ light well upstream (or east)
until $\sim 3''$ (or $\sim 0.9$ kpc) from the H$\alpha$ peak. The three H$\alpha$
peaks downstream are emission-line objects.
The enhanced $B$ band light downstream (relative to $I$ band)
is related to the emission-line objects.
The $I$ band profile at -15$''$ --- -5$''$ can be fitted with an exponential
profile with a scale height of $\sim 6.0''$.
{\bf Middle}: Surface brightness profiles measured parallel to the major axis
of the galaxy (shown as a green box in Fig. 1b, 4$''$ width) in four bands.
The H$\alpha_{\rm off}$ light and the scaled continuum light trace the H$\alpha$
light well beyond the central 5$''$ (or $\sim 1.6$ kpc).
The continuum light within the central 5$''$ is flattened with substructures
(also shown in Fig. 1e and 1f). Surface brightness of some area in the north is
absent because of bright stars there.
If the profiles at -26$''$ --- -6$''$ and 9$''$ --- 37$''$ are fitted with
exponential profiles respectively, the derived scale heights are $\sim 11.9''$
in the south and $\sim 13.3''$ in the north. {\bf Right}:
Surface brightness profiles measured in elliptical annuli are plotted against the
semi-major axis ($a$). The elliptical annuli are centered on the H$\alpha$ peak
of the galaxy and have a fixed axis ratio of 2 ($a/b$, $b$:
semi-minor axis). The positional angle is the same as that used to measure
the surface brightness profiles along the major axis in the middle panel (Fig. 1b).
All unresolved sources (including all emission-line objects but ELO1)
are masked. The $B$ and $I$ profiles at 7$''$ --- 32$''$ can be well fitted with
an exponential profile with a scale height of $\sim 9.5''$ (or 3 kpc).
}
\end{figure}

\end{document}